\documentclass[10pt,letterpaper,twocolumn]{article} 

\usepackage{ol2}
\usepackage[draft]{hyperref}
\usepackage{amsmath}

\newcommand{\micron}{\mu\mbox{m}}
\newcommand{\degrees}{^{\circ}}

\begin{document}

\twocolumn[ 

\title{Fresnel filtering of Gaussian beams in microcavities}


\author{
Susumu Shinohara,$^{1,*}$ Takahisa Harayama,$^2$ and Takehiro Fukushima$^{3}$
}

\address{
$^1$Max-Planck-Institut f\"ur Physik Komplexer Systeme, N\"othnitzer
Stra\ss e 38, D-01187 Dresden, Germany\\
$^2$NTT Communication Science Laboratories, NTT Corporation, 2-4
Hikaridai, Seika-cho, Soraku-gun, Kyoto 619-0237, Japan\\
$^3$Department of Communication Engineering, Okayama Prefectural
University, 111 Kuboki, Soja, Okayama 719-1197, Japan\\
$^*$Corresponding author: susumu@pks.mpg.de
}

\begin{abstract}
We study the output from the modes described by the superposition of
Gaussian beams confined in the quasi-stadium microcavities.
We experimentally observe the deviation from Snell's law in the output
when the incident angle of the Gaussian beam at the cavity interface is
near the critical angle for total internal reflection, providing direct
experimental evidence on the Fresnel filtering.
The theory of the Fresnel filtering for a planar interface qualitatively
reproduces experimental data, and a discussion is given on small
deviation between the measured data and the theory.

\end{abstract}

\ocis{140.2020, 140.3410, 140.4780, 230.3990, 230.5750.}

 ] 

When a beam with finite angular spreads is incident to a dielectric
interface with incident angle near the critical angle for total
internal reflection, it is theoretically shown that its refracted beam
exhibits deviation from Snell's law \cite{Tureci02a,Tureci03}.
Because of the dependence of Fresnel reflection on incident angle,
some angular components are totally internally reflected, while others
far from the critical incident angle are refracted out, resulting in
an angular shift in the far-field emission pattern.
This effect, named the Fresnel filtering (FF) \cite{Tureci02a,Tureci03},
implies the correction of ray optics in addition to the Goos-H\"anchen
effect \cite{Goos47} and has recently attracted attention in
interpreting emission patterns from microcavities
\cite{Rex02,Lee04,Schomerus06,Altmann08,Unterhinninghofen10,Song10}.

In experiments on GaN microlasers with quadrupolar deformed cavities
\cite{Rex02}, emission patterns were explained by the FF effect on the
output from ``scarred modes'', which are associated with the unstable
periodic ray orbit.
In case of the scarred modes, however, general properties on the mode
profile are not well understood, making it difficult to quantitatively
estimate the FF effect.
Considering that the original theory was established for Gaussian
beams \cite{Tureci02a,Tureci03}, it is desirable to experimentally
study the FF effect for the modes that are well described by the
superposition of Gaussian beams.
In this Letter, we study such modes for the quasi-stadium cavities,
where Gaussian mode characteristics can be estimated by the
Gaussian-optic theory \cite{Tureci02b}.
We experimentally demonstrate the deviation from Snell's law and
provide an analysis based on the FF theory.

\begin{figure}[b]
\centerline{\includegraphics[width=6.8cm]{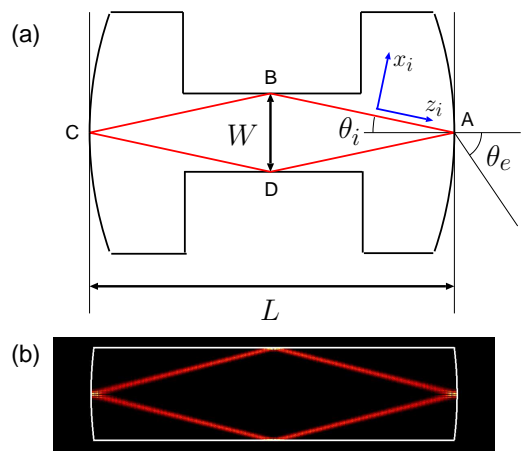}}
\caption{(Color online) (a) The geometry of the quasi-stadium cavity
  with $L=600$ $\micron$ and $W=150$ $\micron$. The radius $R$ of
  curvature of both curved edges is $600$ $\micron$. Inside the
  cavity, the ring orbit reflected at the boundary points A, B, C, and
  D is shown. (b) Resonant mode associated with the ring orbit
  with the wavelength $\lambda\approx 0.856$ $\micron$ calculated by
  the Gaussian-optic theory.}
\label{fig:quasi-stadium}
\end{figure}

Figure \ref{fig:quasi-stadium} (a) shows the geometry of the
quasi-stadium cavity \cite{Fukushima04}.
In this Letter, we study microlasers with quasi-stadium shape with
cavity widths $W$=$100$ $\micron$, $150$ $\micron$, $190$ $\micron$, and
$200$ $\micron$, while we fix the cavity length $L$=$600$ $\micron$.
The radius $R$ of curvature of both curved edges is fixed as $R$=$L$
(i.e., confocal resonator condition).
The devices are fabricated by using a MOCVD-grown gradient-index,
separate-confinement-heterostructure, single-quantum-well
GaAs/Al$_{x}$Ga$_{1-x}$As structure and a reactive-ion-etching
technique \cite{Fukushima04}.

The confocal resonator condition yields that the Fabry-Perot orbit
(bouncing between the boundary points A and C) and the ring orbit
(reflected at A, B, C, and D) are both stable.
In previous studies of the quasi-stadium microlasers, selective
excitation of the Fabry-Perot modes and that of the ring modes were
studied in detail \cite{Fukushima04}.
The selective excitation has been successfully demonstrated by
patterning the ``contact window'', where electric currents are
injected, along the Fabry-Perot orbit or the ring orbit.
In this Letter, we consider the devices with the contact window
patterned along the ring orbit with 5 $\micron$ width (see
Ref. \cite{Fukushima04} for the details on the device structure and
fabrication process) and study output from the ring modes as depicted in
Fig. \ref{fig:quasi-stadium} (b).

In the ray-optic limit, the output can be predicted by Snell's law.
We denote $\theta_i$ and $\theta_e$ the incident and the refracted
angle, respectively, as illustrated in Fig. \ref{fig:quasi-stadium}
(a).
For given $L$ and $W$, the incident angle is given by
\begin{equation}
\theta_i=\sin^{-1}
\left(
\frac{W}{\sqrt{L^2+W^2}}
\right).
\end{equation}

Snell's law implies the refracted angle
$\theta_e$=$\sin^{-1}(n\sin\theta_i)$, where $n$ is the effective
refractive index of the cavity, calculated as $3.3$ from the structure
of the devices.
The ring orbit is confined by total internal reflection when the
incident angle is above the critical angle
$\theta_c=\sin^{-1}(1/n)\approx 17.64\degrees$.
In terms of the cavity geometry, total internal reflection of the ring
orbit occurs for $W> L/\sqrt{n^2-1}\approx 191$ $\micron$.
Below we compare experimentally measured emission patterns with the
predictions by Snell's law.

\begin{figure}[t]
\centerline{\includegraphics[width=6.8cm]{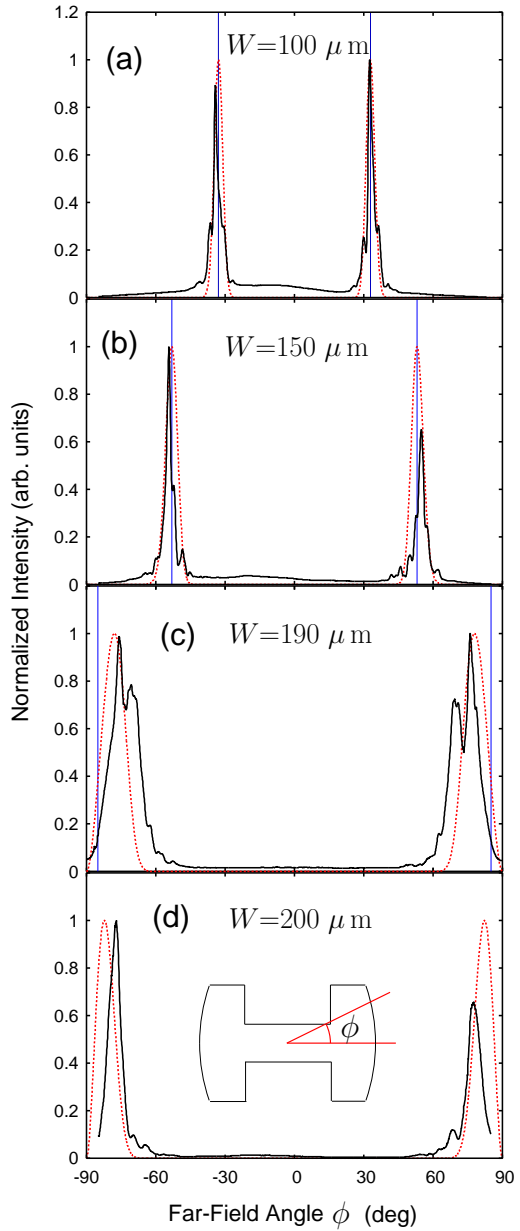}}
\caption{(Color online) Far-field emission patterns of the the
  quasi-stadium microlasers with (a) $W$=$100$ $\micron$, (b)
  $W$=$150$ $\micron$, (c) $W$=$190$ $\micron$, and (d) $W$=$200$
  $\micron$. The definition of the far-field angle $\phi$ is given in
  the inset of (d). Experimental data are plotted with solid curves,
  while the predictions by the FF theory in dotted curves. The
  predictions by Snell's law are indicated by vertical lines.}
\label{fig:ffps}
\end{figure}

In experiments, devices are tested at 25 $\degrees$C using a pulsed
current with 500 ns width at 1 kHz repetition.
The strength of an injection current is determined so that the peak
output power exceeds 5 mW.
Measured far-field patterns for the devices with various cavity widths
are shown in Fig. \ref{fig:ffps} (solid curves).
The definition of the far-field angle $\phi$ is given in
Fig. \ref{fig:ffps} (d).
The patterns are normalized so that the maximum intensity becomes
unity.
We indicate the prediction from Snell's law by vertical lines.

For $W=100$ $\micron$ (i.e., $\triangle\theta=\theta_c-\theta_i\approx
8.2\degrees$) and $W=150$ $\micron$ (i.e., $\triangle\theta\approx
3.6\degrees$), we find that the output peaks are in good agreement
with the predictions from Snell's law.
This agreement convinces us the validity of our estimate of the
effective refractive index $n=3.3$.
For $W=190$ $\micron$ (i.e., $\triangle\theta\approx 0.068\degrees$),
however, we see apparent deviation of the experimental data from
Snell's law.
More strikingly, for $W=200$ $\micron$ (i.e., $\triangle\theta\approx
-0.80\degrees$), Snell's law predicts total internal reflection, but
the experimental data show the output peaks around $\phi=\pm
77\degrees$.

The observed phenomenon is theoretically understood by the FF effect.
Here, we present an analysis based on the theory by Tureci and Stone
\cite{Tureci02a,Tureci03}.
First, we assume that the modes associated with the ring orbit are
well described as the superposition of Gaussian beams.
This is, in fact, justified by the Gaussian-optic theory in the
short-wavelength limit \cite{Tureci02b}.
As the sizes of our cavities are quite large compared to the
wavelength (i.e., $nkR\approx 10^4$), the above assumption is well
satisfied.
Below, we focus on the Gaussian beam
\begin{equation}
E_i(x_i,z_i)=
\frac{E_0 w_0}{w(z_i)}
\exp\left[-\left(\frac{x_i}{w(z_i)}\right)^2+inkz_i\right],
\end{equation}
with $w^2(z_i)=w_0^2-i(2z_i)/(nk)$ scattered at the boundary point A.
The definition of the coordinates $(x_i,z_i)$ is given in
Fig. \ref{fig:quasi-stadium} (a).
The beam waist $w_0$, the distance between the beam waist position and
the interface $z_0$, and wavenumber $k$ can be estimated by applying
the Gaussian-optic theory to the ring orbit.
Letting $(\alpha,\beta)^T$ the eigenvector of the stability matrix for
a segment of the ring orbit (see Ref. \cite{Tureci02b} for details),
we can write $w_0=1/(\sqrt{nk}|\alpha|)$ and
$z_0=-\mbox{Re}(\alpha\beta^*)/|\alpha|^2$.

For a given $W$, we consider the ring mode with the wavelength
$\lambda\approx 0.856$ $\micron$, which corresponds to the lasing
wavelength in the experiments.
In order to simplify the calculation, we employ the Dirichlet boundary
condition, instead of the dielectric boundary condition.
This approximation does not affect so much the modal structure inside
the cavity when the wavelength is sufficiently short.
The field intensity distribution of a calculated mode for $W=150$
$\micron$ is shown in Fig. \ref{fig:quasi-stadium} (b).
From the Gaussian-optic calculation, we found that for the cavities
with $W=100-200$ $\micron$, the Gaussian beam is characterized by
$nk\approx 24$ $\micron^{-1}$ and $w_0\approx 5$ $\micron$, and the
beam waist is located at the boundary points B and D, i.e.,
$z_0=\sqrt{L^2+W^2}/2$.

Next, we consider how the Gaussian beam is refracted at the boundary
point A.
In the polar coordinates $(\rho,\phi)$, the refracted electric field
has the following asymptotic form in the limit $k\rho\to\infty$
\cite{Tureci02a,Tureci03}:
\begin{equation}
E_e(\rho,\phi) \approx
\frac{nk w_0 E_0}{\sqrt{2ik\rho}}
\,t(s_0)\,G(s_0)\,J(\phi,s_0)
\,e^{ik\rho},
\label{eq:E_e}
\end{equation}
where $s_0=s_0(\phi)$ is determined by solving the equation
$n\sin(\theta_i+\delta\theta_i(s))=\sin\phi$ in terms of $s$ with
$\delta\theta_i=\sin^{-1}(s)$.
$t(s)$ is the Fresnel transmission coefficient for the transverse
electric polarization, $G(s)$ carries the information on the incident
Gaussian beam, and $J(\phi,s)$ arises from the stationary phase
approximation in deriving Eq. (\ref{eq:E_e}).
These functions are defined as follows:
\begin{equation}
t(s)=\frac{2n\cos(\Theta_i(s))}
{\cos(\Theta_i(s))+n^2\sqrt{\sin^2\theta_c-\sin^2(\Theta_i(s))}},
\end{equation}
\begin{equation}
G(s)=\exp\left[-\left(\frac{nkw_0}{2}\right)^2 s^2+inkz_0\sqrt{1-s^2}\right],
\end{equation}
\begin{equation}
J(\phi,s)=\frac{\cos\phi\sqrt{1-s^2}}{\sqrt{n^2-\sin^2\phi}},
\end{equation}
where $\Theta_i(s)=\theta_i+\delta\theta_i(s)$.
Using Eq. (\ref{eq:E_e}) with $nk$, $w_0$, and $z_0$ estimated from
the Gaussian-optic theory, we calculated theoretical far-field
patterns, which are plotted with dotted curves in Fig. \ref{fig:ffps}.

When the incident angle $\theta_i$ is far from the critical angle
$\theta_c$ (i.e., for $W=100$ $\micron$ and $W=150$ $\micron$), we find
that the peak positions of the theoretical curves coincide with the
predictions by Snell's law.
However, the FF effect appears when $\theta_i$ is near $\theta_c$.
For $W=190$ $\micron$, we find noticeable deviation between the peak
positions of the theoretical curve and the prediction by Snell's law.
Moreover, for $W=200$ $\micron$, where Snell's law predicts total
internal reflection, we find the maxima of the theoretical curve
around $\phi=\pm 82\degrees$.

When $\theta_i \ll \theta_c$, Eq. (\ref{eq:E_e}) predicts that the
far-field peak position is almost constant with respect to a change of
the beam waist $w_0$, while the peak width varies.
Fitting the experimental data for $W=100$ $\micron$ by
Eq. (\ref{eq:E_e}) with $w_0$ being a free parameter, we found that the
peak width is best reproduced with $w_0\approx 5$ $\micron$.
This convinces us that our estimate of $w_0$ from the Gaussian-optic
theory is reliable.

The theoretical curves qualitatively explain the shifts from Snell's
law observed in the experimental data, but they underestimate the
sizes of the shifts.
The deviation is $4.9\degrees$ for $W=190$ $\micron$, while
$5.0\degrees$ for $W=200$ $\micron$, where for a double peak of $W=190$
$\micron$, we compared the average of the two peak positions with the
peak position of the theoretical curve.

For $\theta_i\approx \theta_c$, Eq. (\ref{eq:E_e}) predicts that the
far-field peak position starts to depend on the beam waist $w_0$.
This suggests a possibility that a slight error in estimating $w_0$
shifts the peak position of Eq. (\ref{eq:E_e}).
For the experimental data for $W=190$ $\micron$ and $W=200$ $\micron$,
we found that in order to explain the measured peak positions by
Eq. (\ref{eq:E_e}), one needs to put $w_0\approx 2.5$ $\micron$, which
deviates too much from the estimate from the Gaussian-optic theory.
Hence, we conclude that the deviations between the FF theory and the
experimental data cannot be solely attributed to errors in $w_0$.

The FF theory assumes an infinite planar interface, whereas the actual
interface has curvature.
This can be another cause for the deviations, although currently we lack
a theory that quantitatively predicts the effect of curvature.
Equation (\ref{eq:E_e}) reproduces the experimental data very well
when $\theta_i\ll\theta_c$ and its deviations from the experimental
data are observed only when $\theta_i\approx\theta_c$.
Therefore, we expect a mechanism making the curvature effect prominent
especially when $\theta_i\approx\theta_c$.

%
%
S.S. acknowledges financial support from DFG research group 760
``Scattering Systems with Complex Dynamics'' and the DFG Emmy Noether
Program.
%

%

\begin{thebibliography}{99}
%
\bibitem{Tureci02a} H. E. Tureci and A. D. Stone,
Opt. Lett. {\bf 27}, 7 (2002).
%
\bibitem{Tureci03} H. E. Tureci, Ph. D thesis, Yale University, 2003.
%
\bibitem{Goos47} F. Goos and H. H\"anchen, Ann. Physik {\bf 436}, 333
  (1947).
%
\bibitem{Rex02} N. B. Rex, H. E. Tureci, H. G. L. Schwefel,
  R. K. Chang, and A. D. Stone, Phys. Rev. Lett. {\bf 88},
  094102 (2002).
%
\bibitem{Lee04} S. -Y. Lee, S. Rim, J. -W. Ryu, T. -Y. Kwon, M. Choi,
  and C. -M. Kim,, Phys. Rev. Lett. {\bf 93},
  164102 (2004).
%
\bibitem{Schomerus06} H. Schomerus and M. Hentschel,
  Phys. Rev. Lett. {\bf 96}, 243903 (2006).
%
\bibitem{Altmann08} E. G. Altmann, G. Del Magno, and M. Hentschel,
  Europhys. Lett. {\bf 84}, 10008 (2008).
%
\bibitem{Unterhinninghofen10} J. Unterhinninghofen and J. Wiersig,
  Phys. Rev. E {\bf 82}, 026202 (2010).
%
\bibitem{Song10} Q. H. Song, L. Ge, A. D. Stone, H. Cao, J. Wiersig,
  J. -B. Shim, J. Unterhinninghofen, W. Fang, and G. S. Solomon,
	Phys. Rev. Lett. {\bf 105},
  103902 (2010).
%
\bibitem{Tureci02b} H. E. Tureci, H. G. L. Schwefel, A. D. Stone, and
  E. E. Narimanov, Opt. Express {\bf 10}, 752 (2002).
%
\bibitem{Fukushima04} T. Fukushima and T. Harayama, IEEE
J. Sel. Top. Quantum Electron. {\bf 10}, 1039 (2004).
%
\end{thebibliography}
\end{document}